# Toy model of harmonic and sum frequency generation in 2D nanostructures

**JIE XU[1], VASSILI SAVINOV[1], AND ERIC PLUM[1,*]**

1 Optoelectronics Research Centre and Centre for Photonic Metamaterials,

University of Southampton, Southampton, SO17 1BJ, United Kingdom

*erp@orc.soton.ac.uk

## Abstract

Optical nonlinearities of matter are often associated with the response of individual atoms. Here, using a toy oscillator model, we show that in the confined geometry of a two-dimensional dielectric nanoparticle a collective nonlinear response of the atomic array can arise from the Coulomb interactions of the bound optical electrons, even if the individual atoms exhibit no nonlinearity. We determine the multipole contributions to the nonlinear response of nanoparticles and demonstrate that the odd order and even order nonlinear electric dipole moments scale with the area and perimeter of the nanoparticle, respectively.

## Introduction

In nonlinear media, electromagnetic waves can interact, generating new waves at combinatorial frequencies. Depending on the generated frequency, the nonlinear phenomena can be classified as harmonic generation, parametric down conversion, sum or difference frequency generation, which are important for optical signal processing, bio-imaging and spectroscopy. Even order harmonics are usually forbidden in materials with inversion symmetry, but with the presence of inhomogeneity associated with interfaces [1-9], optical field gradients [10,11] or chirality [12], centrosymmetric media can support even harmonic response. In particular, nanostructuring can introduce inhomogeneity and asymmetry in metallic and dielectric films, resulting in even order sum-frequency and harmonic generation in structures made from centrosymmetric media [13-22]. Such nonlinearities depend strongly on geometry, i.e. the shape of particles or the unit cell of periodic structures. The advance of nanofabrication to smaller and smaller scales provides an opportunity to engineer the nonlinear properties of particles and (meta)surfaces on the nanoscale. Such engineering depends on understanding the dependence of nonlinear optical properties on the arrangement of atoms. Full ab-initio quantum treatments of surface nonlinearity have been demonstrated, but only for specific materials and under specific conditions [23-29], whilst experimental measurements, of surface nonlinearity, require pristine surfaces and strict control over the bulk effects [3,5,30-32]. Sophisticated classical/quantum test models have been previously developed for metals [33,34], 'dipolium' dielectrics [29,35], and few dielectric particles of quite specific shapes [36]. Meanwhile, symmetry-based selection rules [37,38] identify cases where even harmonic generation is forbidden, but cannot predict its magnitude where it is allowed.

Therefore, the motivation of this work is to develop a simple classical model that can predict how harmonic and sum frequency generation depend on the size and geometry of dielectric nanostructures. We consider a 2D nanoparticle described as a lattice of linear oscillators coupled by Coulomb interactions. The model reveals the different origins of the even and odd order nonlinear processes and predicts the dependence of the efficiency of nonlinear processes on the nanoparticle's geometry, symmetry, size and the incident polarization state. We show that the even order and odd order nonlinear electric dipole moments scale linearly with the perimeter and area of the nanoparticle, respectively, and quantify magnetic dipole and electric quadrupole contributions to nonlinearity. The model may be applied to optimize and understand harmonic and sum frequency generation in dielectric nanostructures of various shapes.

## Results and discussion

### Coulomb-interaction-induced optical nonlinearity of 2D nanoparticles

Previously, using a Taylor series approach, we reported a model of Coulomb-interaction-induced second harmonic generation in the electric dipole approximation [39]. Here, enabled by Fourier transformation, we include higher harmonics, sum frequencies and higher multipoles. Following [39], a dielectric structure is modelled as a collection of atoms, where each atom is described as a classical Lorentz oscillator [40], consisting of one optical electron bound to one nucleus, as shown by Figure 1. The interaction between the electron with coordinate $\boldsymbol{r}(t)$ and its stationary nucleus at $\boldsymbol{R}$ is described by a linear restoring force, which gives rise to a linear optical response. In this model, there is no nonlinearity within a single atom (see Discussion). However, if we consider a particle, which contains $N$ atoms, the Coulomb interaction of electrons with electrons, and electrons with nuclei of other atoms will introduce nonlinearity. The equation of motion for each optical electron can be written as

$$\ddot{\boldsymbol{r}}_k + \gamma\dot{\boldsymbol{r}}_k + \omega_0^2(\boldsymbol{r}_k - \boldsymbol{R}_k) = \frac{q}{m}\boldsymbol{E}(t) + \frac{q^2}{4\pi\epsilon_0 m}\sum_{i\neq k}^{N}\left(\frac{\boldsymbol{r}_k-\boldsymbol{r}_i}{|\boldsymbol{r}_k-\boldsymbol{r}_i|^3} - \frac{\boldsymbol{r}_k-\boldsymbol{R}_i}{|\boldsymbol{r}_k-\boldsymbol{R}_i|^3}\right) \tag{1}$$

where $\boldsymbol{r}_k$ and $\boldsymbol{R}_k$ is the positions of electron $k$ and its nucleus in the particle; $\gamma$ is the damping frequency; $q$ and $m$ are the electron charge and mass; and $\omega_0$ is the angular resonance frequency of an isolated atomic harmonic oscillator.

Second harmonic generation (SHG) due to Coulomb interactions has been observed experimentally in dimers of metallic nanoparticles of deeply sub-wavelength size [41] and a similar nonlinearity arising from electrostatic interactions has been considered in the context of nonlinear plasmonic metamaterials [42]. Similar approaches have been previously employed to study linear [43] and nonlinear properties of dielectrics [44], harmonic generation by metallic structures [45] and anharmonic oscillators [46,47], and for treating surface nonlinearity of uniform planar surfaces [7,27-29,35]. A range of models of sum frequency generation has been reported, e.g. for harmonic oscillators [48] and liquid interfaces [49-51]. Nonlinear oscillator models predicting enhanced nonlinear response of metacrystals have been reported [52,53] and applied to split ring resonator magnetic media. Here our focus is on highly structured materials and our model is applicable to planar dielectric nanostructures of any shape. Therefore, it complements existing models that describe harmonic or sum frequency generation in specific materials and/or shapes [16,23,33,34,36], and symmetry-based selection rules [37,38,54] that predict cases where such processes are forbidden.

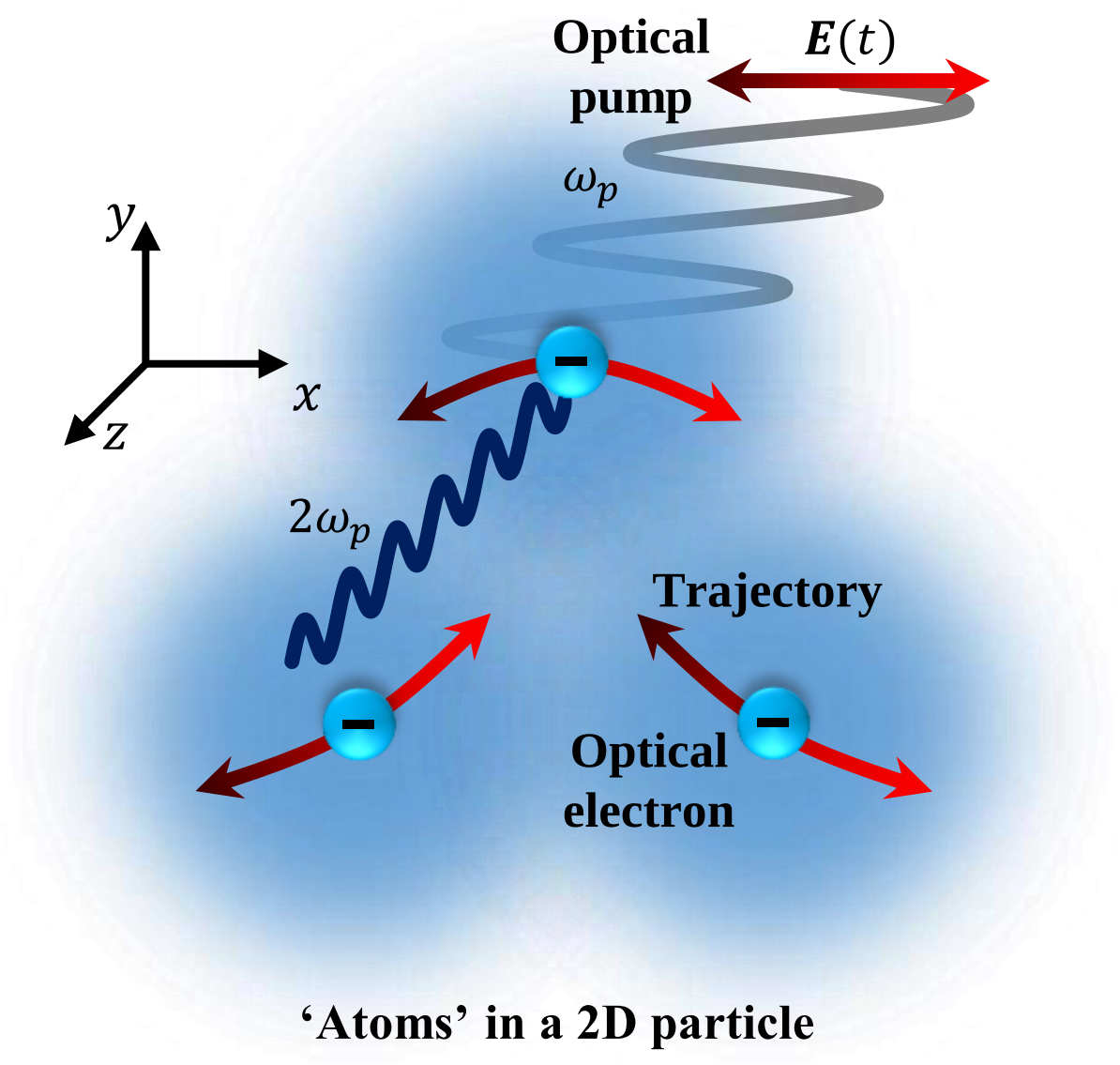

**Fig. 1.** Nonlinear model for the optical response of interacting atoms in a dielectric nanoparticle. The atoms are modelled as damped harmonic oscillators consisting of an optical electron constrained to move in the $xy$-plane and a positively charged stationary nucleus at the atom's centre. The nonlinear optical response of a nanoparticle originates from the Coulomb interactions between optical electrons and other atoms. These interactions cause sum/harmonic frequency generation by perturbing the harmonic electron oscillation driven by incident light field $\boldsymbol{E}(t)$ at the pump frequency $\omega_p$. In particular, Coulomb forces at boundaries become direction-dependent and electron trajectories become curved.

In order to predict the nonlinear response of a structured dielectric film, such as a nanoparticle or metamaterial unit cell, we consider plane wave illumination (along $z$) of a 2D (two-dimensional) lattice of 'atoms' consisting of charges that are confined to the $xy$-plane. The electromagnetic response of the single atom is strictly linear and arises from a harmonic potential [39]. The nonlinear response arises exclusively due to inter-atomic Coulomb interactions. The $\propto 1/\boldsymbol{r}^2$ dependence of the Coulomb force, where $\boldsymbol{r}$ is the inter-charge separation, gives rise to a nonlinear response of optical electrons in collections of coupled atoms, which is illustrated by curved optical electron trajectories in Fig. 1.

We apply this basic principle of one-to-one coupling via the Coulomb force to modelling of the response of large collections of atoms that form particles with a square or hexagonal lattice. The resonance frequency of an isolated atom is $\omega_0 = 9.4\times10^{15}$ rad $s^{-1}$, corresponding to a UV resonance at 200 nm wavelength, and the damping frequency is $\gamma = 0.01\ \omega_0$ to approximate a typical atomic response in dielectrics such as ITO, $TiO_x$ and SiN. In order to access a highly nonlinear regime, the amplitude of the pumping electric field ($E_0$) is set to $8.68\times10^{10}$ V/m, corresponding to $10^{15}$ W/cm$^2$ pump intensity. While this extreme level of intensity may not be feasible experimentally, it allows us to explore physical trends and higher-order optical nonlinearities in a regime where they are of sufficient magnitude to make numerical errors of differential equation solvers irrelevant. The pump wavelength is chosen to be 1064 nm, a common choice for nonlinear optics that is far from the atomic resonance (angular pump frequency $\omega_p = 0.19\ \omega_0$). In modelling of sum frequency generation, we pump the particle at two frequencies, $\omega_p$ and $0.1\omega_p$, simultaneously. The ratio of 10 of the two pump frequencies is chosen to make the sum frequency order easily recognizable. The atomic spacing is chosen to be 0.5 nm in all cases.

By solving the coupled equations of motion [Eq. (1)] of all optical electrons within 2D nanoparticles numerically, we calculate the displacement of each optical electron in the time domain. We solve the coupled system of differential equations in Matlab starting without optical electron displacement and allowing transitional effects to pass before analysing the oscillation of all electrons over 1000 periods of the driving field. This allows the calculation of the particle's net multipole moments. In the interest of simplicity, our initial discussion will be limited to the particle's electric dipole moment in order to describe radiation emitted normal to the plane of the particle. The electric dipole moment is proportional to the displacement of electrons. Therefore, a displacement of the $k^{th}$ electron relative to its nucleus, $\boldsymbol{r}_k - \boldsymbol{R}_k$, results in an electric dipole due to the $k^{th}$ atom of $\boldsymbol{d}_k = q(\boldsymbol{r}_k - \boldsymbol{R}_k)$, and the total electric dipole of the particle is the sum over all atoms, $\boldsymbol{P} = \sum_k \boldsymbol{d}_k$. This gives the electric dipole moment $\boldsymbol{d}(t)$ of each atom and the net electric dipole moment $\boldsymbol{P}(t)$ of the nanostructure in the time domain. The net electric dipole moment is separated into linear and nonlinear components, $\boldsymbol{P}^{(1)}$, $\boldsymbol{P}^{(2)}$, $\boldsymbol{P}^{(3)}$ … , oscillating at the driving frequency $\omega_p$ and its harmonics $2\omega_p$, $3\omega_p$ … , by Fourier transformation. In the presented Fourier series, the peak values are the dipole amplitudes in Coulomb meters at harmonic and sum frequencies. For pumping of a 2D particle at normal incidence, as considered here, radiation emitted normal to the plane of the particle is determined by its net electric dipole moment alone. Towards the end of this paper, we will consider other multipoles, which contribute to radiation in other directions.

## Harmonic and sum-frequency generation in nanostructures of D4 symmetry

Here we consider how a structure in the $xy$-plane radiates along $z$ in response to pumping along $z$. According to symmetry analysis, radiation of even order harmonics along $z$ is forbidden in structures with even-fold rotational symmetry and allowed in structures of no or 3-fold rotational symmetry [37,38]. In order to reveal how nonlinear frequency generation appears/vanishes in cases where it is allowed/forbidden, we studied harmonic and sum frequency generation in simple 2D particles of either square shape (D4 symmetry, with square lattice) or triangular shape (D3 symmetry, with hexagonal lattice).

Figure 2a illustrates harmonic generation in a square particle of 25 atoms. The frequency-dependence of the total electric dipole moment $\boldsymbol{P}$ generated by electric pump field $E_0\cos(\omega_p t)$ is shown, where $P_{i,j}$ refers to the $i$-component of the dipole moment caused by a $j$-polarized pump field, with $i,j$ being $x$ or $y$. As required by symmetry, $x$-polarized excitation generates a dipole only along $x$, while $y$-polarized excitation generates a dipole only along $y$ with the same magnitude. i.e. $P_{x,x}=P_{y,y}$ are observed, while $P_{x,y}$ and $P_{y,x}$ are forbidden. As expected for a particle with even-fold rotational symmetry [37,38], we observe only odd harmonics – $\boldsymbol{P}^{(1)}$, $\boldsymbol{P}^{(3)}$, $\boldsymbol{P}^{(5)}$, $\boldsymbol{P}^{(7)}$ and $\boldsymbol{P}^{(9)}$ – which appear as peaks in Fig. 2a.

Harmonic generation is a special case of sum frequency generation. The latter becomes apparent when pumping the square particle with a combination of two frequencies, $\omega_p$ and $0.1\omega_p$. This is illustrated by Figure 2b, which considers a pump field $E_0/2[\cos(\omega_p t)+\cos(0.1\omega_p t)]$. As for harmonic generation, also mixed frequency pumping results in an electric dipole moment of the particle that is parallel to the pump polarization, $P_{x,x}$ and $P_{y,y}$. We observe only odd order sum frequencies, i.e. odd harmonics of the higher and lower pump frequencies ($\omega_p$, $3\omega_p$, $5\omega_p$, … and $0.1\omega_p$, $0.3\omega_p$, …) and sums of odd combinations of both pump frequencies ($1.2\omega_p$, $2.1\omega_p$, $2.3\omega_p$ …), indicating that the model can predict wave mixing in 2D particles. We observe that different dipole moments of the same order tend to be of similar magnitude, but resonant enhancement near the atomic resonance at 5.26 $\omega_p$ and the background level also influence the magnitude of the calculated dipole moments. (Similar results are obtained for different choices of pump frequency, see Supplementary Figure S1.)

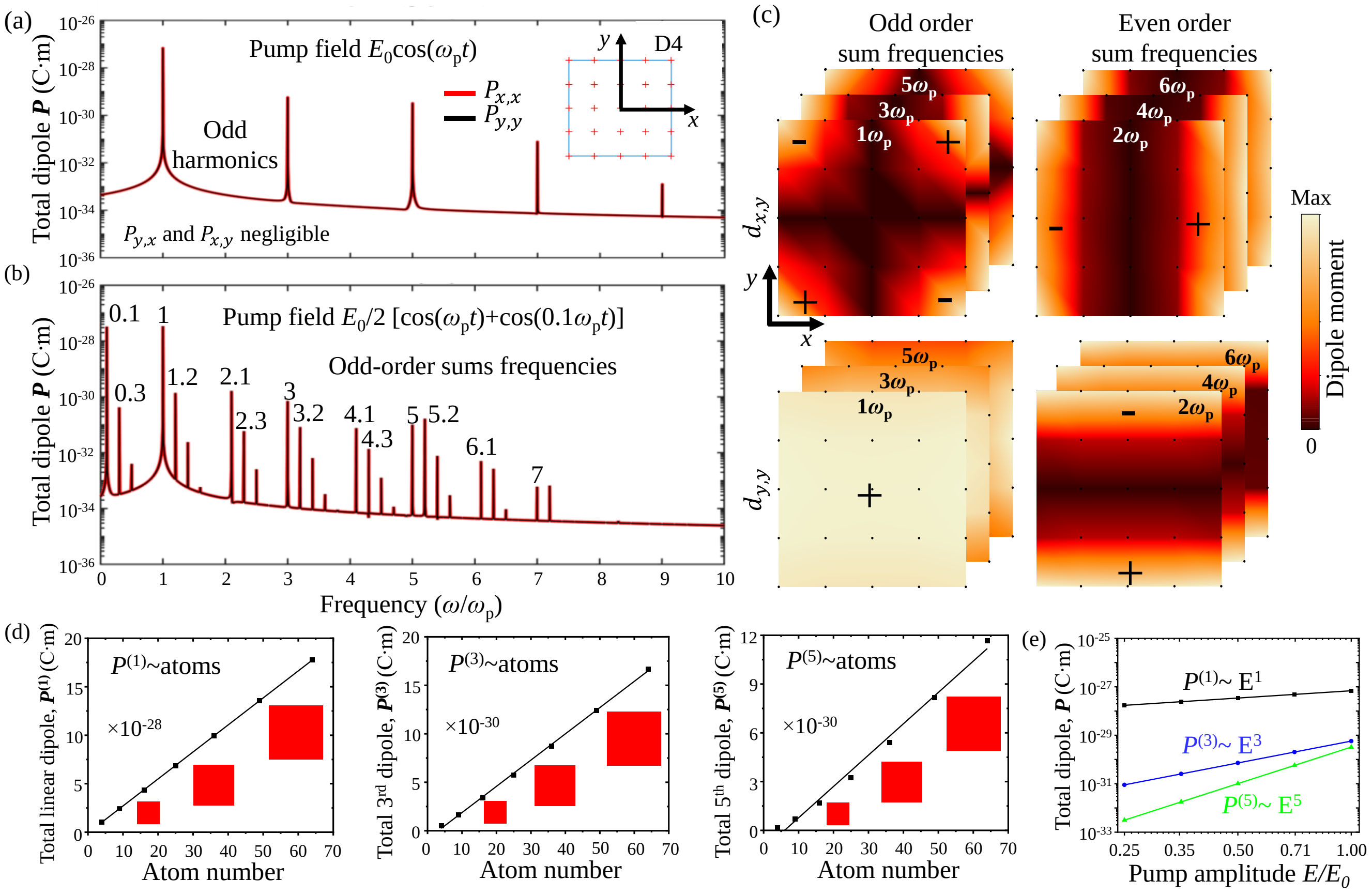

**Fig. 2.** Harmonic and sum frequency generation in a nanoparticle of D4 symmetry. (a)-(b) Frequency dependence of the electric dipole moment of a square particle (blue outline) consisting of 25 atoms (red crosses) in response to optical pumping at (a) a single frequency $\omega_p$ and (b) a combination of two frequencies, $\omega_p$ and $0.1\omega_p$. $P_{y,x}$ indicates the $y$-component of the particle's electric dipole moment caused by $x$-polarized pumping. For either $x$- or $y$-polarized pumping, the orientations of pump polarization and generated dipole are parallel, and the generated dipoles have the same magnitude for both cases. (c) Magnitude (colours) and sign ("+" and "-") of the dipole moment per atom, $\boldsymbol{d}$ generated at harmonic frequencies (multiples of $\omega_p$) in response to pumping at frequency $\omega_p$. The top (bottom) row shows the dipole component orthogonal (parallel) to the pump polarization. Stacked images for different – either even or odd – harmonics show the same qualitative behaviour. (d) Scaling of the particle's electric dipole moment with its size at the first three odd harmonic frequencies [corresponding to peaks in (a)]. These odd order dipole moments scale with the particle's number of atoms (area). (e) Scaling of the particle's electric dipole moment with the pump field amplitude for the same cases.

The origin of harmonic and sum frequency generation is revealed by the distribution of atomic dipole moments at the relevant frequencies. Figure 2c illustrates this for the square particle pumped at frequency $\omega_p$. The top row shows the distributions of dipole components oriented normal to the pump polarization at odd (left) and even (right) harmonic frequencies. Notably, all harmonics are generated locally, but the symmetry of the particle implies that the atomic electric dipole component normal to the pump field has the same magnitude and a $\pi$ phase difference on opposite sides of the structure, resulting in overall cancellation. Therefore, the structure of D4 symmetry has no net electric dipole moment in the direction normal to the pump polarization. The net electric dipole component parallel to the pump polarization also cancels for even harmonics (bottom row, right). In contrast, the dipole component parallel to the pump polarization does not vanish for odd harmonics (bottom row, left). In-phase oscillation of all atoms causes the atomic dipoles to add up to a net electric dipole moment of the particle at odd harmonic frequencies.

Indeed, the symmetry of the experimental arrangement, including structure and pump beam, means that no light can be emitted at even harmonics along $z$ and therefore the net electric dipole moment must vanish. In these cases, the fields "radiated" from different parts of the particle cancel by destructive interference in the normal direction. Radiation in other directions may not vanish, can be described in terms of other multipoles, and will be discussed towards the end of this paper. Here, we describe radiation emitted normal to the plane of the particle.

While the allowed component of odd harmonic generation originates from the entire particle, we notice that the even harmonics – which do not add up to any net electric dipole moment – are associated with edges of the particle. In the interest of simplicity, we have shown dipole distributions for harmonic frequency generation in Fig. 2c. Dipole distributions for sum frequency generation with different pump frequencies (see Supplementary Figure S2) have the same characteristics as those for harmonic generation of the same order.

The dependence of the nanoparticle's odd harmonic response on its size is shown by Fig. 2d. We find that the total nonlinear electric dipole at the generated odd harmonics is proportional to the total number of atoms (2D particle area). The pump power dependence of the generated harmonics is illustrated by Fig. 2e, which shows the electric dipole moment generated in the square particle for different pump field amplitudes at different harmonic frequencies. The magnitude of the electric dipole moments at the first, third and fifth harmonic frequencies – $\boldsymbol{P}^{(1)}$, $\boldsymbol{P}^{(3)}$, $\boldsymbol{P}^{(5)}$ – is proportional to the first, third and fifth power of the pump amplitude, as expected for first, third and fifth order (non)linear optical processes.

## Harmonic and sum-frequency generation in nanostructures of D3 symmetry

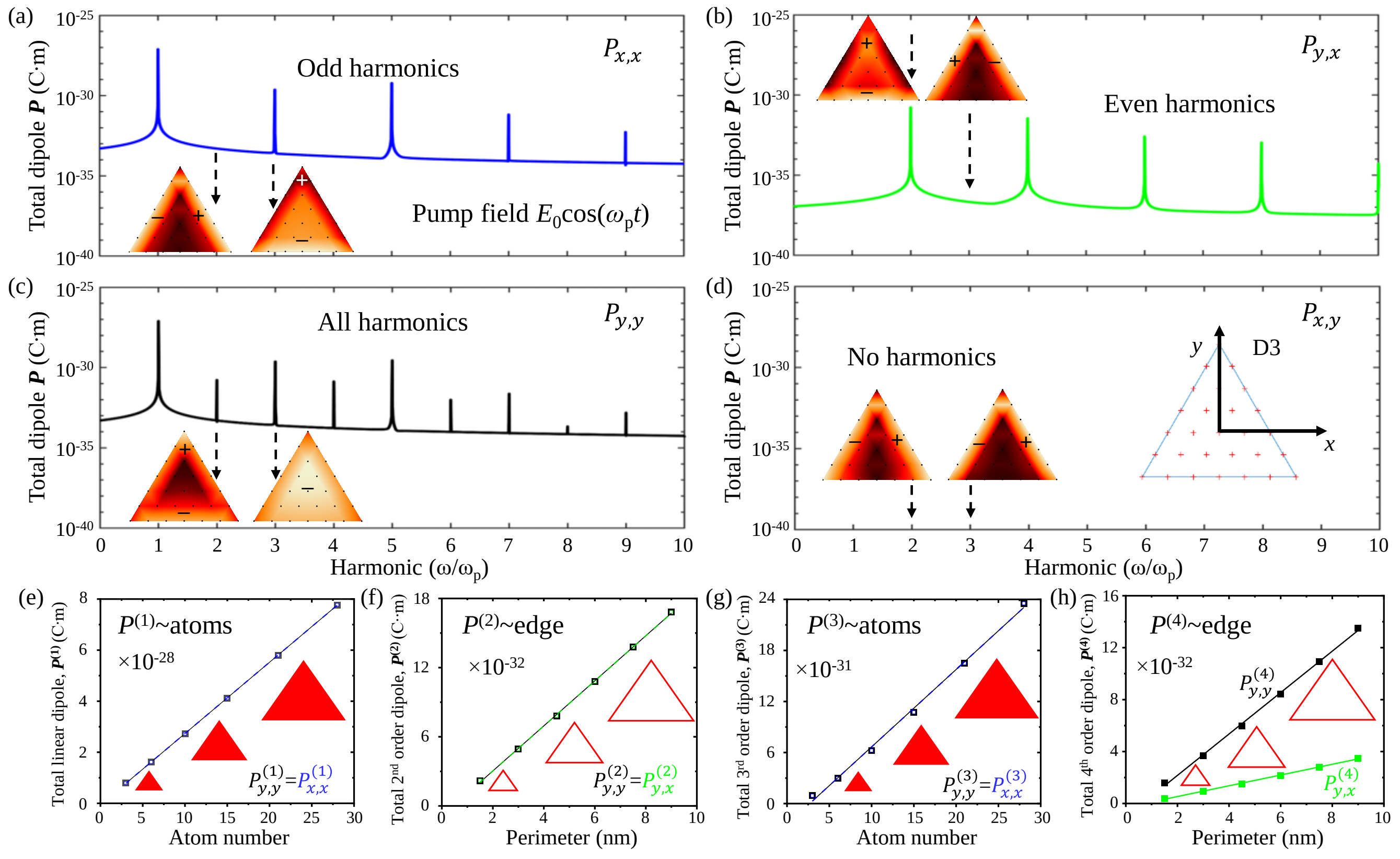

**Fig. 3.** Harmonic generation in a nanoparticle of D3 symmetry. (a)-(d) Frequency-dependence of the electric dipole moment of a triangular particle pumped at frequency $\omega_p$. Components of the dipole moment (a, c) parallel and (b, d) orthogonal to the (a, b) *x*-polarized and (c, d) *y*-polarized pump field. Absence of curves in case (d) indicates cancellation of the electric dipole component at all frequencies. The inset shows the outline (blue line) and the 28 atoms (red crosses) of the triangular particle. Colourmaps show the distribution of the atomic dipole moments at the 2nd and 3rd harmonic frequencies, which are representative of even and odd harmonics. Colour represents the dipole magnitude and regions oscillating in anti-phase are indicated by opposite signs as in Fig. 2c. (e-h) Scaling of the particle's dipole moment with its size at the first four harmonic frequencies. The particle's electric dipole moment scales with its number of atoms (area) at odd harmonics and with its perimeter length at even harmonics.

While the even-fold rotational symmetry discussed above only permits odd order nonlinear electric dipole moments, three-fold rotational symmetry allows a more complex nonlinear response. Now we will discuss harmonic generation in a particle of D3 symmetry, a triangular particle consisting of 28 atoms arranged in a hexagonal lattice as shown by Figure 3. The spectral dependence of the triangular particle's electric dipole moment in response to pumping at frequency $\omega_p$ is shown by Fig. 3a-d. Also here, the electric dipole moment at the $k^{th}$ harmonic is proportional to the $k^{th}$ power of the pump field amplitude, causing higher harmonics to vanish quickly with reducing pump amplitude, as it should be (see Supplementary Figure S3). An *x*-polarized pump field generates an electric dipole moment of the particle along *x* at odd harmonic frequencies (Fig. 3a) and along y at even harmonic frequencies (Fig. 3b). A *y*-polarized pump field generates a dipole moment along *y* only at all harmonic frequencies (Fig. 3c,d). The distribution of the atomic dipole moments within the nanoparticle is shown for the second and third harmonic frequencies (insets), which are representative of the even and odd harmonic cases. While the particle exhibits local charge oscillations along $x$ and $y$ at all harmonics, overall cancellation of components of the particle's electric dipole moment at even (Fig. 3a), odd (Fig. 3b) or all (Fig. 3d) harmonics results from charge oscillations with equal amplitude and opposite phase at opposite edges of the triangle.

It has been reported that the total linear and second-order nonlinear responses of a particle scale differently with its size. For a 3D particle made from centrosymmetric material, the total linear electric dipole moment is proportional to its volume (number of atoms), while the second-order dipole is proportional to its surface area [30]. For a 2D particle, as considered here, the linear electric dipole moment is proportional to its area (number of atoms), while the second-order electric dipole is proportional to its perimeter [39]. However, how does particle size affect higher order harmonics? Fig. 3e-3h show how the electric dipole moment of a triangular 2D particle changes with particle size. The magnitudes of the total dipole moments, $\boldsymbol{P}^{(1)}$, $\boldsymbol{P}^{(2)}$, $\boldsymbol{P}^{(3)}$, $\boldsymbol{P}^{(4)}$, at the first four harmonic frequencies are shown for triangular particles with perimeter from 1.5 nm (3 atoms in total) to 9 nm (28 atoms). We find that the total nonlinear electric dipole at odd harmonics is proportional to the total number of atoms (2D particle area), while the dipole at even harmonics is proportional to the particle's perimeter. At the first three harmonic frequencies, $x$-polarized and $y$-polarized pump light induce dipole moments of the same magnitude, while at the fourth harmonic $P_{y,y}^{(4)} = 4P_{y,x}^{(4)}$.

Sum frequency generation in the triangular particle is consistent with harmonic generation. An $x$-polarized pump electric field at a combination of two different frequencies generates a nonlinear electric dipole moment along $x$ for odd sum frequencies and along $y$ for even sum frequencies, while a $y$-polarized pump field generates a nonlinear dipole moment along $y$ only for all sum frequencies, where more and more sum frequencies are generated with increasing pump amplitude (Supplementary Figure S4). For sum frequency generation resulting from cross-polarized pump fields of different frequencies see Supplementary Figure S5. In this case, $x$-polarized electric dipole moments are generated at sum frequencies containing an odd multiple (1, 3, …) of the $x$-polarized pump frequency and generated $y$-polarized sum frequencies contain an even multiple (0, 2, …) of the $x$-polarized pump frequency.

## Multipole contributions to harmonic generation

While radiation normal to the plane of the 2D particle pumped at normal incidence is determined by its electric dipole moment, other multipoles[55] contribute to radiation in other directions.

The $\alpha$-component of the particle's net electric dipole moment is given by

$$P_\alpha = \int d^3\mathrm{r}\, \rho(\mathbf{r})\, \mathrm{r}_\alpha = -q \sum_{i=1}^{N} r_{i,\alpha} \tag{2}$$

where $\rho(\mathbf{r})$ is the charge density as a function of position $\mathbf{r}$. In our 2D case, only $P_x$ and $P_y$ are allowed and the radiated power is given by the following expression, where $Z_0$ is the impedance of free space, $c$ is the speed of light and $k = \omega/c$

$$\mathcal{P}^{(P)} = \frac{c^2 Z_0}{12\pi} k^4 \|\boldsymbol{P}\|_2^2 = \frac{c^2 Z_0}{12\pi} k^4 \left(|P_x|^2 + |P_y|^2\right) \tag{3}$$

The $\alpha$-component of the particle's magnetic dipole moment is

$$m_\alpha = \frac{1}{2c} \int d^3\mathrm{r}\, [\mathbf{r} \times \boldsymbol{J}(\mathbf{r})]_\alpha = \frac{-q}{2c} \sum_{i=1}^{N} [(\boldsymbol{R}_i + \boldsymbol{r}_i) \times \dot{\boldsymbol{r}}_i]_\alpha \tag{4}$$

where $\boldsymbol{J}$ is the current density and $\dot{\boldsymbol{r}}_i$ is the electron velocity. Here, only $m_z$ is allowed and the radiated power is

$$\mathcal{P}^{(m)} = \frac{c^2 Z_0}{12\pi} k^4 \|\boldsymbol{m}\|_2^2 = \frac{c^2 Z_0}{12\pi} k^4 |m_z|^2 \tag{5}$$

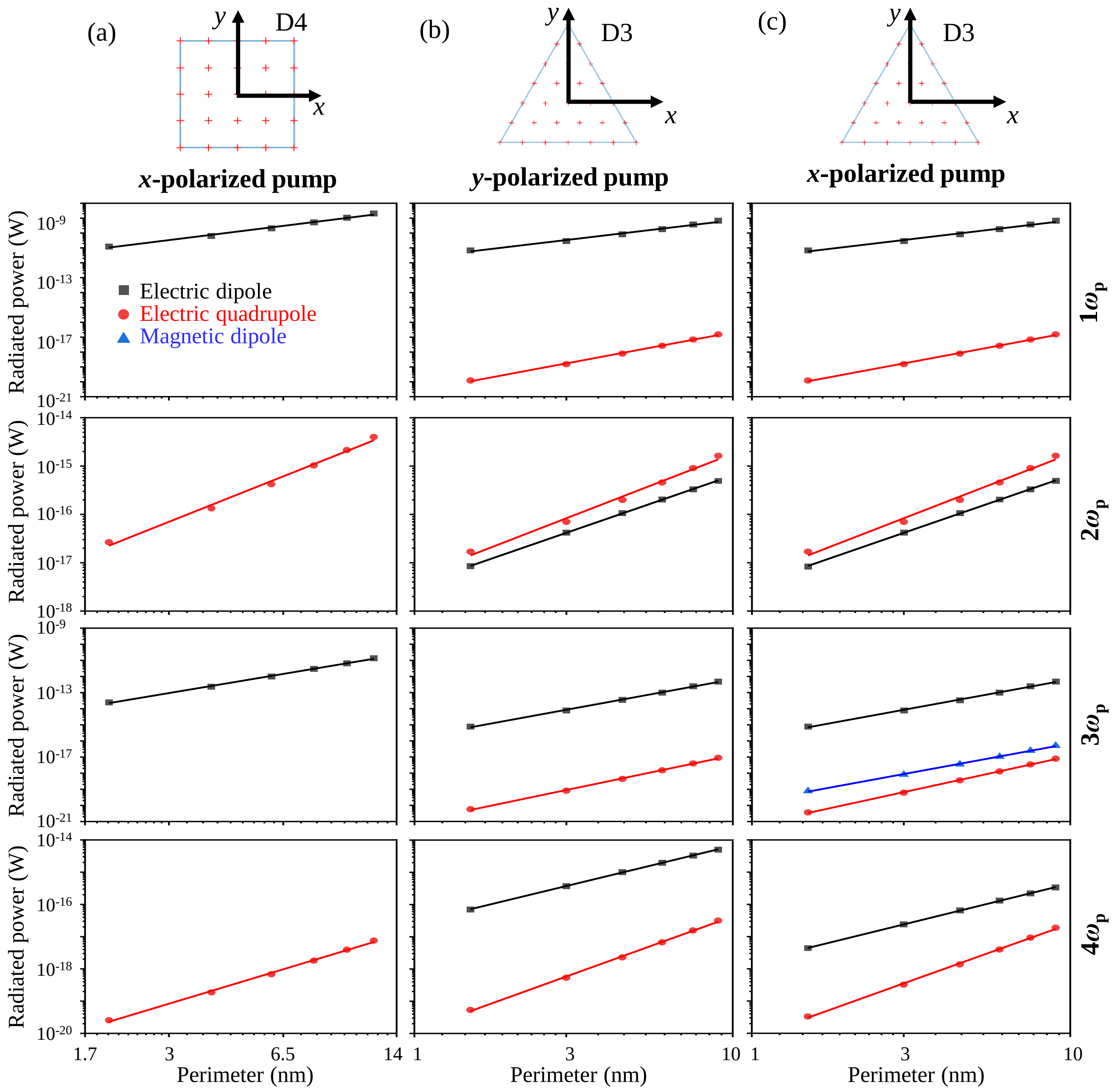

**Fig. 4.** Power radiated by different multipoles at harmonic frequencies for particles of (a) D4 and (b, c) D3 symmetry as a function of particle perimeter. Only the electric dipole (black) radiates along *z*.

The $\alpha, \beta$-component of the particle's electric quadrupole is given by

$$Q_{\alpha,\beta} = \frac{1}{2}\int d^3\mathfrak{r}\,\rho(\mathbf{r})\left[\mathfrak{r}_\alpha \mathfrak{r}_\beta - \frac{1}{3}\delta_{\alpha,\beta}\mathfrak{r}^2\right] = \frac{q}{2}\sum\nolimits_{i=1}^{N}\left[R_{i,\alpha}R_{i,\beta} - \frac{1}{3}\delta_{\alpha,\beta}\boldsymbol{R}_i^2 - (R_{i,\alpha}+r_{i,\alpha})(R_{i,\beta}+r_{i,\beta}) + \frac{1}{3}\delta_{\alpha,\beta}(\boldsymbol{R}_i+\boldsymbol{r}_i)^2\right]$$
$$= \frac{q}{2}\sum\nolimits_{i=1}^{N}\left[-R_{i,\alpha}r_{i,\beta} - R_{i,\beta}r_{i,\alpha} - r_{i,\alpha}r_{i,\beta} + \frac{1}{3}\delta_{\alpha,\beta}(2\boldsymbol{R}_i\boldsymbol{r}_i + \boldsymbol{r}_i^2)\right] \tag{6}$$

where $\delta_{\alpha,\beta}$ is the Kronecker delta. In our geometry, $Q_{x,x}$, $Q_{y,y}$ and $Q_{x,y} = Q_{y,x}$ are allowed and the radiated power is

$$\mathcal{P}^{(Q)} = \frac{c^2 Z_0}{40\pi} k^6 \sum_{\alpha,\beta} \left|Q_{\alpha,\beta}\right|^2 = \frac{c^2 Z_0}{40\pi} k^6 \left(2\left|Q_{x,y}\right|^2 + \left|Q_{x,x}\right|^2 + \left|Q_{y,y}\right|^2\right) \quad (7)$$

Fig. 4 shows the power radiated by different multipoles for the square and triangular particles, where the coordinate origin is placed at the centre of each particle. For the square particle, electric dipole radiation is emitted at odd harmonics and considerably weaker electric quadrupole radiation ( $Q_{x,x}$ , $Q_{y,y}$ ) is emitted at even harmonics, while magnetic dipole contributions are negligible, Fig. 4a. Therefore, only odd harmonics will be radiated along z. In other directions, all harmonics will be radiated, with even harmonics being several orders of magnitude weaker than odd harmonics of similar order. For the triangular particle, electric dipole radiation dominates at odd harmonics, while electric dipole and quadrupole radiation have a comparable magnitude at the second harmonic frequency, Fig. 4b,c.

The multipole contributions are also revealed by the polarization surface densities (electric dipole distributions) in Fig. 2c and Fig. 3a-d. For example, for both the square and triangular particles, the net quadrupole at even harmonics originates from electrons on opposite edges of the particle, that oscillate towards and away from each other (→←,←→). The net magnetic dipole at the third harmonic frequency for $x$-polarized pumping of the triangular particle arises from anti-parallel electron oscillations (↑↓,↓↑), see Fig. 3b.

## Discussion

The presented model describes the contribution of Coulomb interactions between different atoms to the nonlinear response of dielectric nanostructures. To focus on the universal collective nonlinear response of any dielectric, we purposefully remove all possible atom-specific nonlinearities by describing atoms with the harmonic potential of a classical Lorentz oscillator model (and pumping far away from resonance). However, real atoms can be nonlinear. Considering the inversion symmetry of atoms in the ground-state, opposite directions of electron displacement yield opposite restoring forces. Therefore, the simplest approximation of a nonlinear atom would involve a symmetric anharmonic potential. Within this approximation, an individual atom can only have odd orders of optical response, i.e. contribute to 3$^{rd}$, 5$^{th}$, … order nonlinearity. With respect to the particle's electric dipole, such atomic nonlinearity would contribute to the odd-order components that are parallel to the excitation field. This atomic contribution to a particle's odd order nonlinear dipole moment will be proportional to the particle's number of atoms, just like the contribution from Coulomb interactions between atoms (Fig. 2d, Fig. 3g). Therefore, the atomic nonlinearity will affect the magnitude of the particle's odd order nonlinear electric dipole response parallel to the pump field, but not its qualitative behaviour. It has no bearing on the particle's even order nonlinearity, which arises entirely from interactions between atoms.

## Conclusion

In summary, we have shown that harmonic and sum frequency generation in 2D dielectric nanostructures can be modelled in terms of harmonic oscillators (atoms) coupled by the Coulomb force. This approach allows modelling of large collections of atoms and mapping of the origin of harmonic and sum frequency generation in nanostructures. Odd order sum frequency generation has been found to originate from the whole structure with corresponding net electric dipole moments scaling with the number of atoms. Even order sum frequency generation has been traced to the boundaries of 2D nanostructures, with corresponding net electric dipole moments scaling with the structure's perimeter. The model allows the design and optimization of nonlinear dielectric nanostructures for nanophotonics, such as structures and particles cut from 2D materials and dielectric layers that are thin compared to the pump wavelength. 3D geometries may also be considered by solving Equation (1) in three dimensions.

## Acknowledgements

The authors thank Nikolay Zheludev for helpful advice. This work is supported by the UK's Engineering and Physical Sciences Research Council (grants EP/M009122/1 and EP/T02643X/1), and the China Scholarship Council (CSC No.201706310145).

## Author contributions statement

J.X. conducted the simulations. J.X. and E.P. wrote the manuscript. All authors analysed the results. E.P. and V.S. supervised the project.

## Data availability statement

After a period of embargo, the data from this paper will be available from the University of Southampton ePrints research repository: https://doi.org/10.5258/SOTON/D1792

## Supplemental document

See Supplement 1 for supporting content.

# Supplementary Information

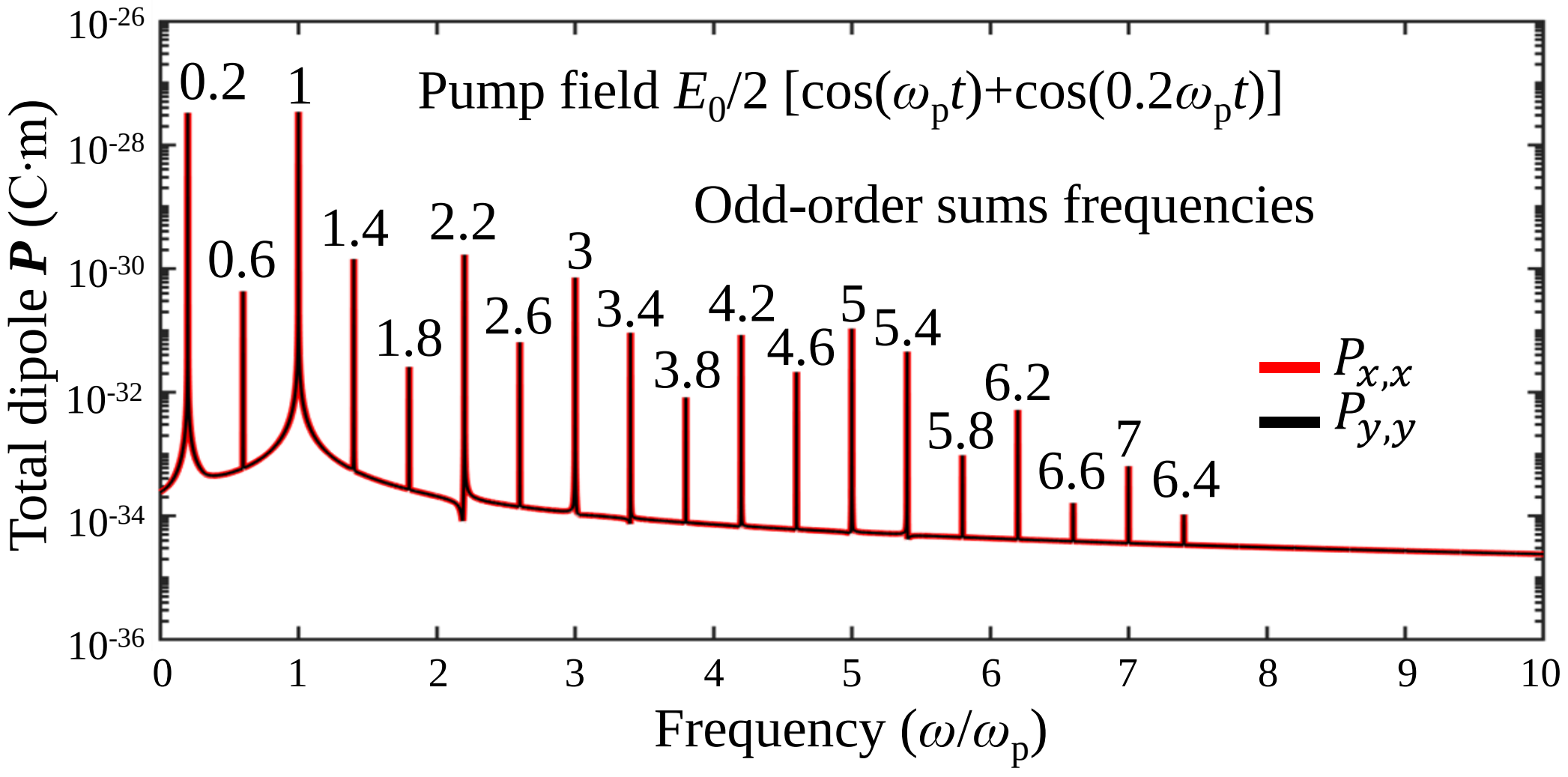

**Fig. S1.** Sum frequency generation in a nanoparticle of D4 symmetry. Frequency dependence of the electric dipole moment of a square particle consisting of 25 atoms in response to optical pumping at a combination of two frequencies, $\omega_p$ and $0.2\omega_p$. $P_{y,x}$ indicates the $y$-component of the particle's electric dipole moment caused by $x$-polarized pumping. For either $x$- or $y$-polarized pumping, the orientations of pump polarization and generated dipole are parallel, and the generated dipoles have the same magnitude for both cases.

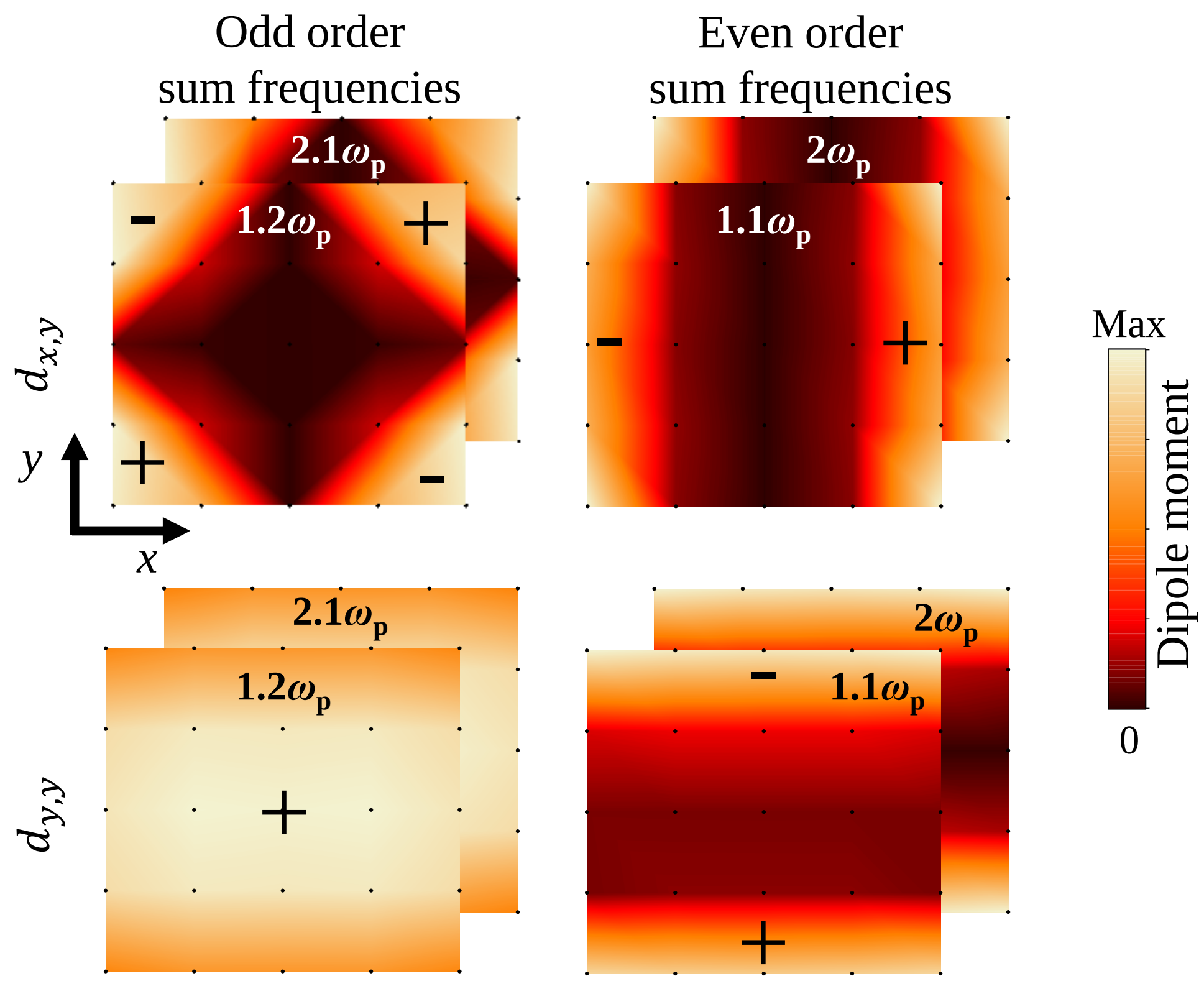

**Fig. S2.** Magnitude (colours) and sign ("+" and "-") of the dipole moment per atom, $\boldsymbol{d}$, generated at lowest-order sum frequencies in response to pumping at frequencies $\omega_p$ and $0.1\omega_p$. The top (bottom) row shows the dipole component orthogonal (parallel) to the pump polarization. Stacked images for different – either even or odd – sum frequencies show the same qualitative behaviour.

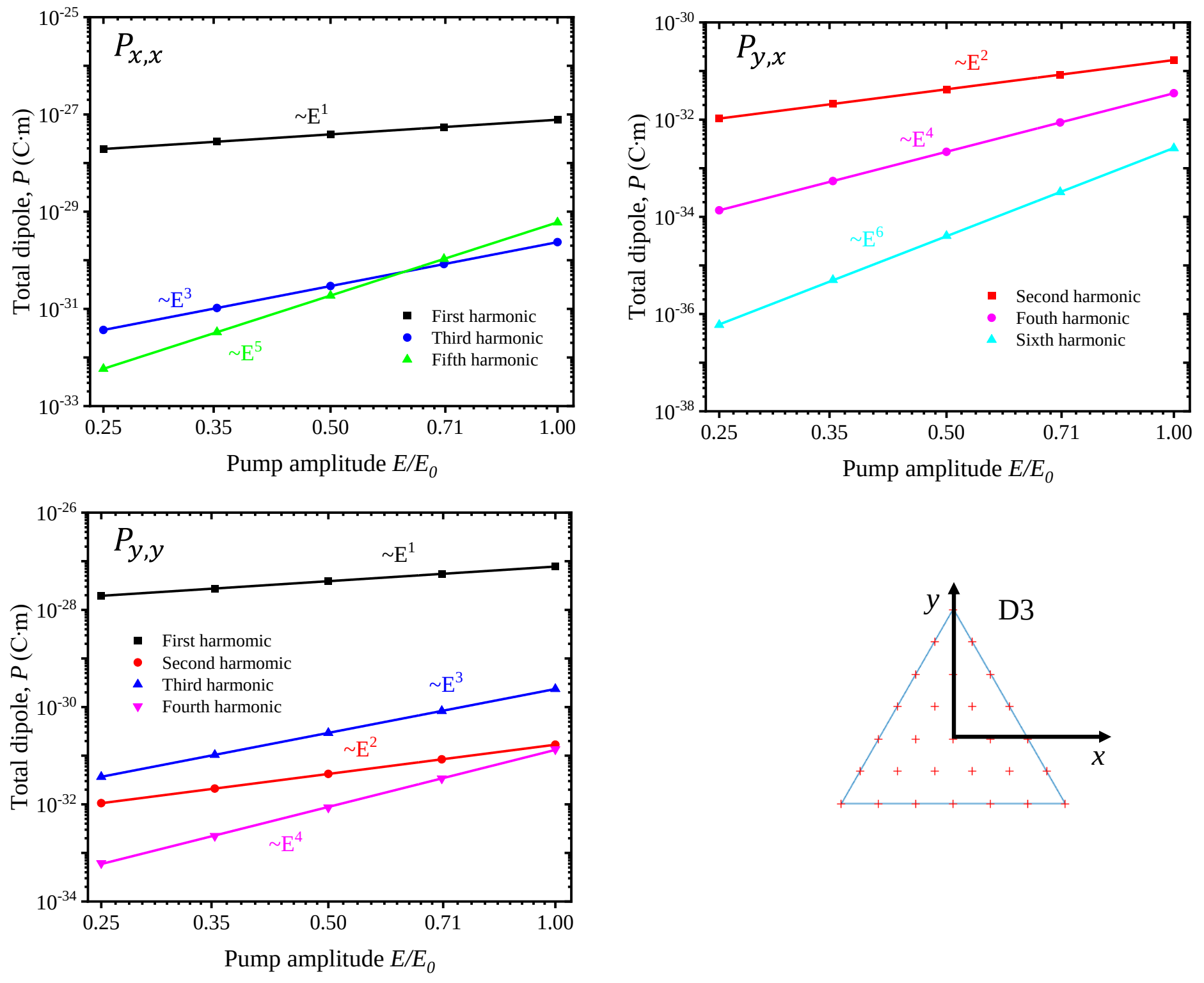

**Fig. S3.** Scaling of the triangular particle's electric dipole moment at harmonic frequencies with the pump field amplitude for the cases of Fig. 3.

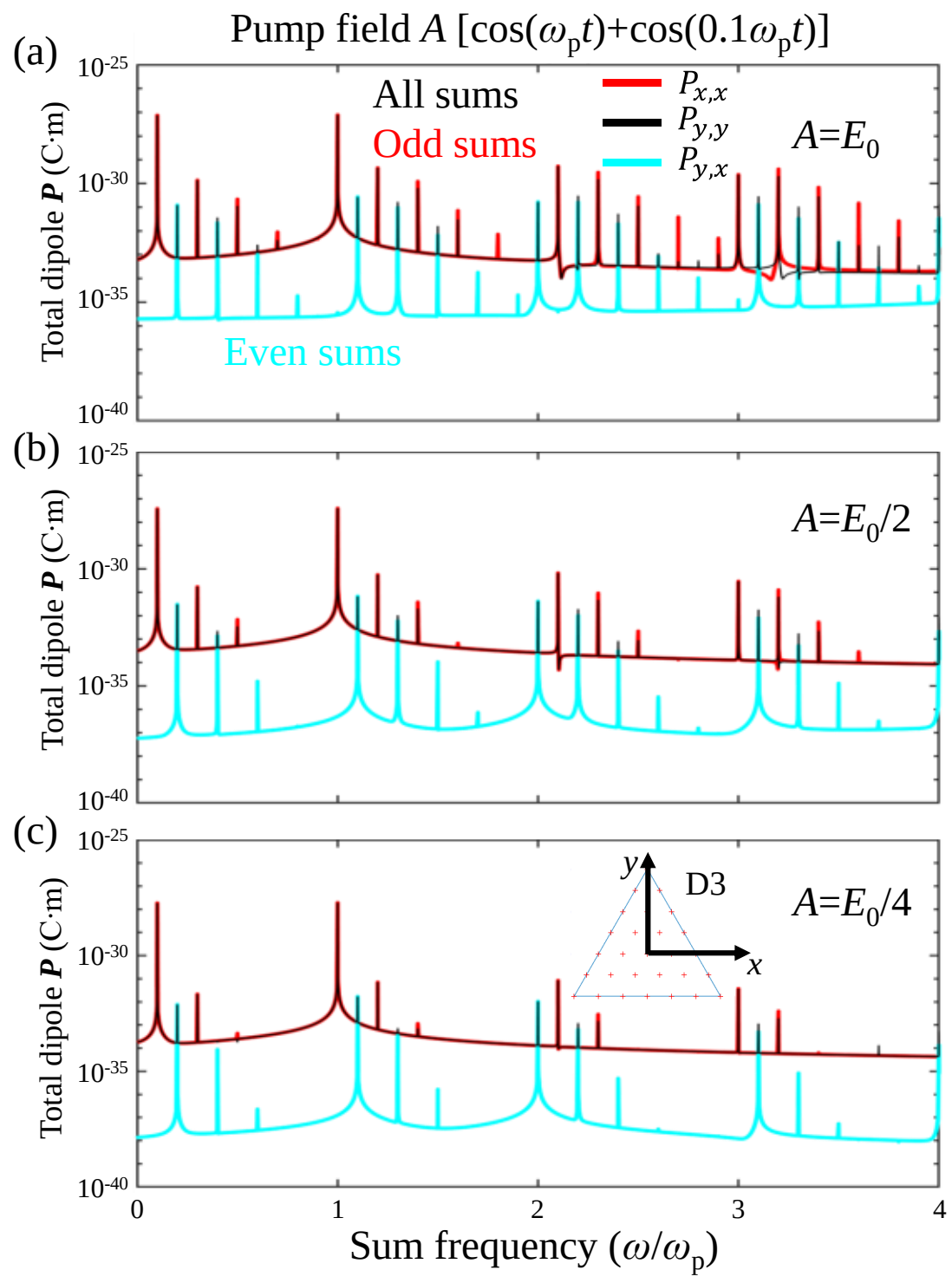

**Fig. S4.** Sum frequency generation in a structure of D3 symmetry with co-polarized pumping. (a)-(c) Frequency dependence of the electric dipole moment of the triangular particle of 28 atoms (inset) in response to pumping at a combination of two frequencies, $\omega_p$ and $0.1\omega_p$ for different pump electric field amplitudes of (a) $E_0$, (b) $E_0/2$, (c)$E_0/4$.

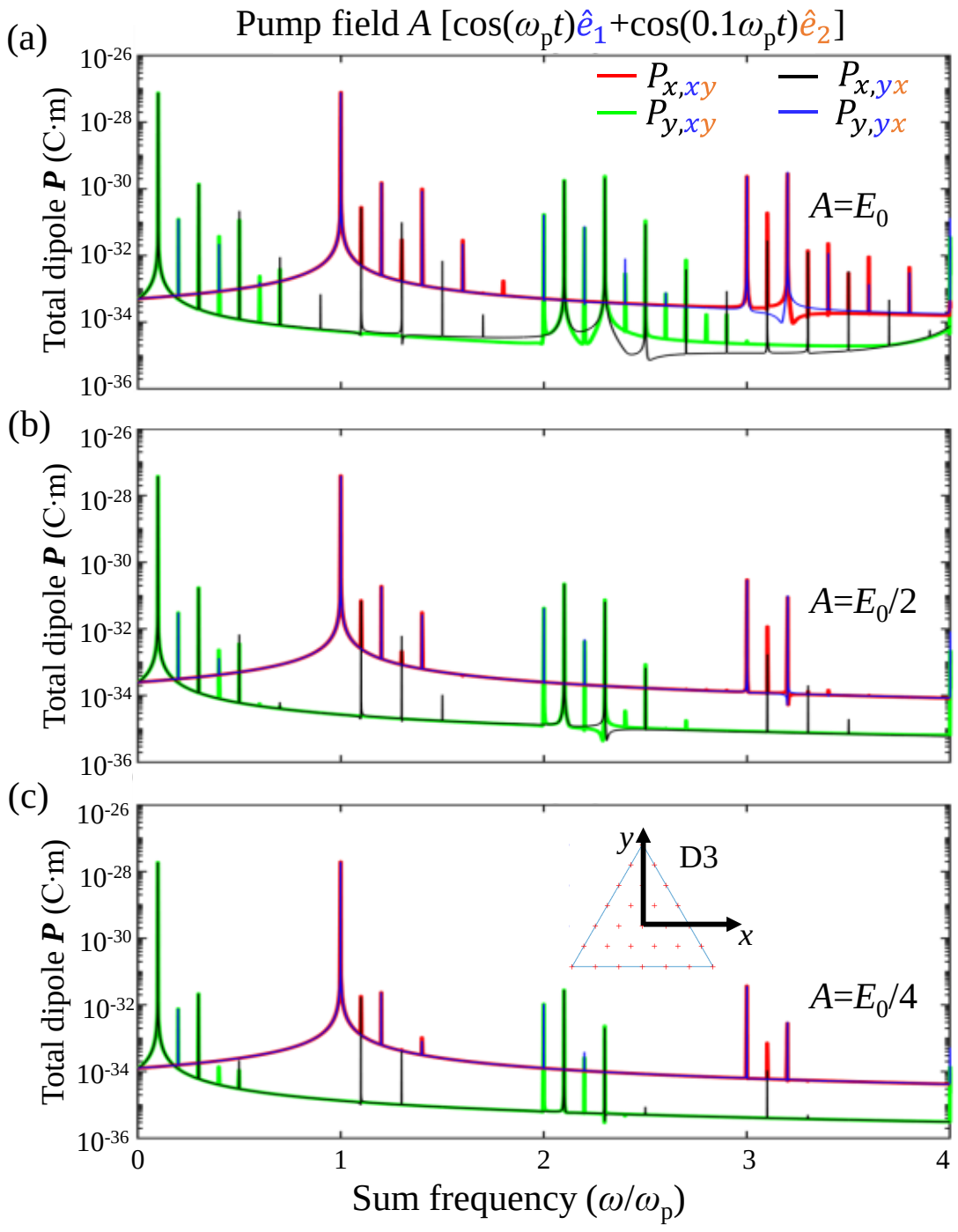

**Fig. S5.** Sum frequency generation in a structure of D3 symmetry with cross-polarized pumping. (a)-(c) Frequency dependence of the electric dipole moment of the triangular particle (inset) in response to pumping at two frequencies, $\omega_p$ and $0.1\omega_p$, where the pump fields at different frequencies have orthogonal polarizations. $P_{i,jk}$ refers to the *i*-component of the dipole moment caused by *j*-polarized pump field at $\omega_p$ and *k*-polarized pump field at $0.1\omega_p$, where *i*,*j*,*k* is *x* or *y*. Different panels show different pump electric field amplitudes of (a) $E_0$, (b) $E_0/2$, (c)$E_0/4$.